# SEMICONDUCTOR HIGH-ENERGY RADIATION SCINTILLATION DETECTOR

A. Kastalsky[a], S. Luryi[a,*], and B. Spivak[b]

[a]University at Stony Brook, ECE Department and NY State Center for Advanced Sensor Technology, Stony Brook, NY 11794-2350

[b]Department of Physics, University of Washington, Seattle, WA 98195

**Abstract**

We propose a new scintillation-type detector in which high-energy radiation produces electron-hole pairs in a direct-gap semiconductor material that subsequently recombine producing infrared light to be registered by a photo-detector. The key issue is how to make the semiconductor essentially transparent to its own infrared light, so that photons generated deep inside the semiconductor could reach its surface without tangible attenuation. We discuss two ways to accomplish this, one based on doping the semiconductor with shallow impurities of one polarity type, preferably donors, the other by heterostructure bandgap engineering. The proposed semiconductor scintillator combines the best properties of currently existing radiation detectors and can be used for both simple radiation monitoring, like a Geiger counter, and for high-resolution spectrography of the high-energy radiation. An important advantage of the proposed detector is its fast response time, about 1 ns, essentially limited only by the recombination time of minority carriers. Notably, the fast response comes without any degradation in brightness. When the scintillator is implemented in a qualified semiconductor material (such as InP or GaAs), the photo-detector and associated circuits can be epitaxially integrated on the scintillator slab and the structure can be stacked-up to achieve virtually any desired absorption capability.

## 1. Introduction

There are two large groups of solid-state radiation detectors, which dominate the area of ionizing radiation measurements, scintillation detectors and semiconductor diodes. The scintillators detect high-energy radiation through generation of light which is subsequently registered by a photo-detector, typically a photo-multiplier that converts light into an electrical signal. The main advantage of existing scintillators is their large detection volume. Semiconductor diodes employ reverse biased p-n junctions or metal-semiconductor junctions where the absorbed radiation creates electrons and holes, which are separated by the junction field thereby producing a direct electrical response. The sensitivity of diode detectors depends on the length of the field region. To increase this length, the doping level in the field region must be minimized. At present, the semiconductor diode is best for the spectral resolution of the ionizing radiation.

As reviewed extensively by Knoll [1], both groups of detectors have their drawbacks, resulting in a lower than desired signal response and resolution. The diodes typically suffer from inadequate electron-hole collection, i.e. not every electron-hole pair created by the radiation results in a current flow in the measurement circuit. The most common semiconductor materials used for the radiation detectors are Si and Ge p-n junctions, where the intrinsic carrier concentration can be reduced to a very low level, while the excellent material properties provide for good electric field uniformity. In the case of silicon, an additional procedure of Li doping is commonly applied to neutralize acceptors in the depletion region, in order to obtain an acceptable junction depletion length. The Si:Li detectors, however, need low temperatures, both during the operation and in storage. Both Si and Ge radiation detectors require relatively high voltages, typically of order kilovolts, to maximize the collection of electrons and holes and increase their drift velocity. This results in an additional unwelcome noise in the current response and leads to problems of surface conductance and voltage breakdown. Even at these high voltages, the response time is larger than 100 ns, limited by the saturated electron and hole drift velocities at high fields. Finally, the dependence of the shape of the output pulse rise on the position at which the electron-hole pairs are created, significantly complicate the measurements.

In the case of scintillators, the efficiency of converting the high-energy radiation into light is typically about 10% (12% in NaI). The reason for this is fundamental: the scintillator material must be transparent to the radiation it produces. To accomplish this, the wide-gap material (7 eV for NaI) is activated with impurities such as thallium which represent recombination sites for electrons and holes. Thus produced light has much lower energy (3 eV for Tl) than the bandgap of the host crystal, whence the poor efficiency. In addition, the recombination time on such impurities is several hundreds of nanoseconds (e.g., 230 ns for NaI activated with Tl), which is undesirably long for fast timing or high counting rate applications. Finally, the high bandgap inherent in all commercially available scintillators implies a relatively high energy (25 eV for NaI) required per each electron-hole pair created by the ionizing radiation. This reduces the detector resolution.

A group at the Lawrence Berkeley National Laboratory has been working on a semiconductor scintillator wherein a direct-gap semiconductor like CdS or ZnO is doped with donors in a reducing atmosphere to provide electrons in shallow states below the conduction band [2], and with radiative centers to trap ionization holes [3, 4]. If the radiative centers are ionized acceptors, self absorption can be reduced by limiting their concentration. If the radiative centers are isoelectronic atoms, holes are trapped in local states, and the subsequent lattice relaxation results in a Stokes shift that further reduces self absorption. The success of these approaches depends on the elimination of non-radiative centers that limit luminosity [5].

We propose a new scintillation-type semiconductor detector in which high-energy radiation produces electron-hole pairs in a direct-gap semiconductor material that subsequently undergo *interband recombination,* producing infrared light to be registered by a photo-detector. The key issue is how to make the semiconductor essentially transparent to its own infrared light, so that photons generated deep inside the semiconductor slab could reach its surface without tangible attenuation. We contemplate two ways to accomplish this. One (relatively inexpensive) way, based on heavy doping of bulk semiconductor with shallow impurities of one polarity type, preferably donors, is discussed in Sect. 2. The allowable slab thickness depends on temperature and is ultimately limited by free-carrier absorption to about 1 mm (Sect. 3). This limitation is essentially lifted in our other approach, discussed in Sects. 4 and 5, which requires an epitaxially grown thick heterostructure of variable bandgap. This approach is, naturally, more expensive to produce, but the additional slab thickness it offers, especially in room-temperature operation, should justify the effort. Both devices are contemplated for the implementation in compound semiconductor materials, such as GaAs or InP, where a mature optoelectronic technology exists. This enables a novel system architecture, discussed in Sect. 6, where each relatively thin (say, 1 mm thick) semiconductor slab is supplied with its own, epitaxially grown or grafted on the surface, photo-detector system. Such systems can then be stacked up without limit, thus increasing the active detector volume to accommodate large absorption length of high-energy radiation.

## 2. Scintillator based on Burstein shift in bulk semiconductor

The key requirement for a scintillator is to be transparent to its own radiation, so that the photons produced deep inside the material can reach the surface and be collected. We propose that this requirement can be fulfilled on the basis of the so-called Burstein shift of the absorption edge [6] in semiconductors doped with impurities *of one type.*[1]

To maximize the internal light emission efficiency, the following material requirements must be fulfilled:

1. The material must be chosen in such a way that the radiative component of recombination dominates over non-radiative components.

2. The material structure must minimize the absorption of its own radiation.

This list of requirements leads to the well-known direct-gap III-V semiconductors, such as GaAs and InP. We shall present our discussion in the instance of InP:

    a. The material is direct, and can provide high internal emission efficiency with the predominantly radiative component of recombination;

    b. Indium has a relatively high atomic number Z=49 (vs. 14 for Si and 32 for Ge), while InP has a relatively low energy of ~ 4 eV per electron-hole pair created by the primary ionizing radiation;

---

[1] When semiconductor is degenerately doped, the edge of absorption is blue-shifted relative to the emission edge by the carrier Fermi energy. This effect underlies the operation of semiconductor lasers.

    c. The effective electron mass in InP is relatively small (0.08 $m_0$) which leads to a lower conduction band density of states and therefore higher Burstein shift for a given doping level.

The proposed detector comprises a slab of direct-gap semiconductor, such as InP, heavily doped so as to minimize the absorption coefficient for its own interband radiation, as illustrated in Fig. 1. Doping with donors is preferable, because electrons typically have a lower effective mass than holes, producing a higher Fermi energy $E_F$ and larger Burstein shift for the same level of doping. The effect exponentially depends on the Fermi level, in that the absorption mean free-path $\lambda$ is exponentially increased: $\lambda = \lambda_0 \exp(E_F/kT)$, where $\lambda_0$ is the interband absorption mean free-path in the undoped material ($\lambda_0 \sim 1$ μm in InP).

For example, for the donor concentration $N_D = 10^{19}$ cm$^{-3}$ in InP, the ratio $E_F/kT \sim 8$ at room temperature, thus yielding the absorption mean free path $\lambda \sim 3$ mm. Further increase of the electron density, desirable for the increase of the $E_F/kT$ ratio, becomes impractical due to the rise of the free-carrier absorption, limiting the useful device thickness, as further discussed below.

Besides making the semiconductor transparent to the emitted photons, the heavy doping shortens the radiative recombination time $\tau$ of minority carriers, according to $\tau = 1/BN_D$, where $B = 10^{-10}$ cm$^3$/s is the radiative recombination coefficient. For the above mentioned doping level of $10^{19}$ cm$^{-3}$ in the InP one has $\tau \sim 10^{-9}$s. Typically, the non-radiative time is in the range of $\sim 10^{-7}$s. Therefore, the ratio $\xi$ of the radiative recombination time to the non-radiative time is about $\xi \sim 0.01$. Thus, the device experiences practically no losses through the non-radiative channel of recombination while still being the fastest high-energy radiation detector ever, with the response time of 1 ns.

The dominance of the radiative recombination channel allows us to increase the active region of the detector beyond the absorption mean-free-path $\lambda$. Indeed, the process of self-absorption by itself does not imply loss of photons and attenuation of scintillating photoemission. Self-absorption creates new electrons and holes and is followed by their recombination and hence re-generation of scintillating photons. The direction of propagation of the re-generated photons is not correlated with the direction of initial absorbed photon. This means that when the thickness $L$ of the semiconductor slab constituting the active zone is thicker than $\lambda$, the propagation of scintillating photons is a diffusive process. In such a process the distance covered is proportional to the square root of the number of steps $N_{\text{step}}$. We can estimate this number as: $N_{\text{step}} \sim (L/\lambda)^2$.

For many applications of the detector it is important to keep $\xi N_{\text{step}} \ll 1$ and the ultimate limitation of the extent of the active zone arises from this requirement. Indeed, consider a slab which is ten times thicker than the mean free path, $\alpha L \sim 10$. This means the expected number of diffusion steps is $N_{\text{step}} \sim 100$. Bearing in mind that a fraction of photons $\xi = 0.01$ is lost at each step to non-radiative processes, we find that after $N_{\text{step}}$ only a fraction $\exp(-\xi N_{\text{step}}) = 37\%$ of photons remain in the flux. For some applications this may be acceptable, but having in mind quantifiable energy resolution, we can anticipate significant analysis problems arising from the fact that the yield of scintillating photons to the output surface becomes dependent on the position in the slab, where the initial electron-hole pairs were generated by ionizing radiation.

For energy-resolution applications we must choose the slab thickness *L* so that the value of $\xi N_{\text{step}}$ is much less than unity, exemplarily in the range of ~ 0.1. In this case, the internal collection efficiency exp(-$\xi N_{\text{step}}$)~0.9, and the yield is approximately independent of the initial ionization position. For $\xi$~0.01 the number of diffusion steps should therefore not exceed, say, $N_{\text{step}}$ ~16. At room temperature, $N_{\text{step}}$ ~ 16 corresponds to the slab thickness $L = (N_{\text{step}})^{1/2} \times \lambda$ ~ 4 × 3 mm = 1.2 cm. It should be noted, however, that the response time will also increase due to the recombination delay at each step. For our example of InP doped to $10^{19}$ cm$^{-3}$, the diffusive light propagation with 16 steps will result in the response time of $N_{\text{step}}\tau$ =16 ns.

From this example, it is clear that the non-radiative rate is an important parameter for our detector and it should be kept to the minimum. We should be concerned, for example, that non-radiative processes may be enhanced at heavy doping. Although in general shallow impurities are not active as non-radiative recombination centers, the process of heavy doping may be accompanied by a simultaneous introduction of parasitic impurities and defects that may act as "lifetime killers". If, for example, the non-radiative time were to decrease to $10^{-8}$s (so that $\xi$= 0.1), then one needs $N_{\text{step}}$< 2 to maintain the internal collection efficiency at the level of ~0.9. In this case, it is preferable to rely on a single photon pass across the device active area and keep the device thickness L~ $\lambda$.

In the "cleanly" doped InP or GaAs semiconductor (no parasitic lifetime killers) the dominant nonradiative process is the Auger recombination, see, e.g., [7]. For Auger-limited recombination, one can write $\xi = CN_D / B$, where *C* is the Auger coefficient. The values of *C* are similar in both InP and GaAs, $C \approx 10^{-30}$ cm$^6 / s$ at room temperature, but the value of the radiative coefficient is substantially larger in GaAs, $B \approx 7 \cdot 10^{-10}$ cm$^3 / s$. This may become a design consideration for choosing the material. According to the experimental data [7], for $N_D = 10^{19}$ cm$^{-3}$ in InP, the relative rate of Auger processes corresponds to $\xi$ ~ 0.01. For GaAs the value of $\xi$ is still lower.

## 3. Free-carrier absorption

It is clear that to operate our detector we must keep the device thickness below the free-carrier absorption length. Free-carrier absorption (FCA) coefficient in semiconductors is not a universal parameter because absorption of photons by free electrons is forbidden by the conservation of momentum and energy. The FCA coefficient thus essentially depends on a third agent, impurity or phonon. The FCA in InP was studied by Dumke and coworkers [8]. It turns out that for photon energies below but near the fundamental bandgap (the region we are interested in) the dominant absorption process is due to free-carrier transitions to higher-lying conduction-band valleys, see Fig. 2. For the free-carrier concentration of $10^{19}$ cm$^{-3}$ in InP, the absorption coefficient is about 7 to 10 cm$^{-1}$, corresponding to $\lambda \approx$ 1 mm. Similar conclusions were reached earlier by Balslev [9] for GaAs. The measured free-carrier absorption coefficient scales rather accurately with the carrier concentration.

Figure 2 illustrates different optical absorption processes in a direct-gap semiconductor near the band-edge, $h\nu \approx E_G$. The interband absorption (process 1) is suppressed unless $h\nu > E_G + E_F$. The intraband free-carrier absorption (2) dominates at longer wavelengths but for $h\nu \approx E_G$ it is about 10 times less efficient than process (3) due to free-carrier transitions between different conduction bands. In fact, from the free-carrier absorption data Dumke et al [8] were able to estimate the separation $\Delta E$ to the satellite (X) valley at 0.92 eV, close to the modern value $\Delta E \approx 0.85$ eV.

The fact that free carrier absorption scales linearly with the carrier concentration means that we should be able to make the slab thicker than 1 mm by going to cryogenic temperatures and using the fact that the efficiency of Burstein shift exponentially increases with decreasing temperature. At cryogenic temperatures the suppression of interband absorption is so effective that the material transparency for the interband photons becomes entirely dependent on free carrier absorption. Figure 3 shows the calculated absorption length λ (mean-free-path with respect to interband absorption) as a function of temperature. One can see a rapid rise of the mean-free-path as the temperature decreases. This rise is limited, however, by free-carrier absorption which is shown as a slant horizontal line for the light wavelength of 0.92 μm.

The ultimate limitation is, therefore, provided by the intersection of the concentration dependence of free-carrier absorption with the interband absorption dependence. Thus determined maximum slab thickness $d_{max}$ is evidently temperature dependent. To a very good approximation, we find $d_{max} \propto 1/T$.

A promising way of circumventing the free-carrier absorption limitation at room temperature is discussed in Sect. 5.

## 4. Photodetector

The detector of light emitted by the scintillator can be integrated with the InP based structure, e.g., implemented as an epitaxially grown interdigitated planar fast photo-receiver (ternary or quaternary InGaAsP) at one the active volume surfaces. This configuration can have special advantages for high-energy physics experiments since such a detector will not be particularly sensitive to external magnetic field *B* and therefore can be used in a high *B* chamber. Furthermore, the photoreceiver structure can be pixellated, thus providing imaging capability of the high-energy particle source. Even more importantly, the integrated detector enables three-dimensional integration, to be discussed in Sect. 6.

Alternatively, the light detector can be located outside the active semiconductor body, e.g. as a photomultiplier tube (PMT). In this case means for enhancing the transfer of light from InP body to the photo-receiver must be engineered. To minimize internal reflections at the InP/photoreceiver interface, a fluid with a high refraction index can be introduced. Such fluids are commercially available with the refraction index of up to 2.5. Also, the output surface can be appropriately textured. Texture techniques for enhancing the emission of light from semiconductor to air have been perfected by the light-emitting diode (LED) community [10]. Recent results in LED technology show that an extraction efficiency as high as ~60% can be achieved [11]. In this case, a photo-multiplier with an infrared photo-cathode is the photoreceiver of choice. All other surfaces of the InP crystal except that facing the photoreceiver should be coated with a metallic film providing mirror reflection of light signal. At this time, the appropriate mirror coating is envisioned as a 50 nm Al film on a thin $SiO_2$ intermediate layer.

**5. Heterostructure scintillator**

As discussed in Sect. 3, the uniform scintillator, based on the Burstein shift is limited by free-carrier absorption to a thickness of about 1 mm at room temperature. Let us now discuss a modified structure, where free-carrier absorption is largely eliminated, Fig. 4.

The epitaxial structure comprises two alternating materials, e.g. $InP/Ga_{0.47}In_{0.53}As$ or $Al_xGa_{1-x}As/GaAs$, lattice-matched to each other. The materials are assumed to have different energy gaps, $E_{G1} = E_{C1} - E_{V1}$ and $E_{G2} = E_{C2} - E_{V2}$, respectively, with the second material having the lower bandgap, $E_{G1} > E_{G2}$. We further assume the second material may be doped, while the first material is largely undoped. From the standpoint of minimizing the dark currents in optical detector at room temperature it makes sense not to make the second bandgap unnecessarily narrow. For this purpose, it may be preferable to use lattice-matched InP/InGaAsP combination with the quaternary material bandgap of about 1.1 eV.

The essential idea is that the total volume occupied by the second material is small compared to that occupied by the first material. For example, if a 2μm-thick InP layers are alternated by a 20 nm-thick layers of InGaAs, the layer thickness ratio is 100 (duty cycle factor δ=0.01).

Upon interaction with the ionizing radiation, the created electrons and holes quickly, within about a nanosecond, diffuse to the InGaAs wells and recombine there. The difference in the band-gap energies (1.35eV for InP and 1.1 eV for InGaAsP) guarantees that all light emission occurs in the InGaAs wells, so that the wider-gap InP remains substantially transparent to the emitted photons.

The only remaining absorption in the heterostructure comes owing to self-absorption in the wells. This effect of interband self-absorption can be further suppressed by increasing the carrier density in the wells, effecting the Burstein shift, as discussed above in Sect. 2. The free-carrier absorption in the wells remains, but is reduced by the duty cycle factor. With the doping density in the wells $N_D=10^{19}$ $cm^{-3}$ and δ=0.01, we obtain the total effective absorption coefficient of less than 0.1 $cm^{-1}$. Thus, the active material in the heterostructure detector is practically transparent to the emitted light at room temperature. This allows one to produce large active detector volume in all three directions, which is essential for efficient absorption of gamma-radiation. High electron density in the narrow-gap wells guarantees both that the radiation recombination is a dominant process and that the response time is about 1 nanosecond.

It should be emphasized that the Burstein shift remains important for the viability of our heterostructure detector. The recombination wells can be viewed as artificial "giant traps" for electrons and holes, which act essentially as efficient radiative recombination centers without introducing non-radiative recombination. Without the Burstein shift, the efficient radiative recombination would translate into equally efficient absorption. In order to achieve a centimeter absorption length the wells would have to be spaced out by a large distance. Suppression of the absorption in the wells by the Burstein shift circumvents this essential problem.

The described heterostructure scintillator can be made several cm thick and is capable of efficient and fast operation even at room temperature. However, the fabrication of this device is not simple. Indeed, most epitaxial growth techniques capable of nano-resolution, such as the metalorganic vapor phase epitaxy (MOCVD) [12], are associated with a relatively slow growth process. The challenge is entirely in the growth time that may span several days, but because of the superior material quality produced with MOCVD, it may be still preferred and used advantageously. The other preferred method is the hydride vapor phase epitaxy, HVPE, which allows a growth rate of 100 μm per hour and even higher. Presently HVPE is primarily used for growing free standing GaN structures [13].

We envision the epitaxial growth of 300 μm-thick films that may be free standing, i.e. grown on a thin substrate that is subsequently removed. By making a stack of such films, we obtain the desired detector volume. Exemplarily, 33 heterostructure films are stacked up in the growth direction to make a 1 cm thick detector structure. To minimize the internal light reflections, an intermediate film comprising a transparent fluid, powder or epoxy having a high index of refraction (ideally, close to that of InP) can be placed between the films.

## 6. Three-dimensional integration

An important advantage of the proposed scintillator concept is that it can be implemented within a highly developed semiconductor technology, based on GaAs or InP, both widely used in modern electronics and photonics. This means that each detector slab can be endowed not only with a photo-receiver array but also with the counting and decision making circuitry. Therefore, we can contemplate a three-dimensional integration of scintillator-detector systems representing semiconductor wafers or chips of standard thickness (0.5 mm) to achieve virtually unlimited thickness in the integrated system.

One can envision three different levels of a three-dimensional detector integration, Fig. 5. In the simplest case, shown in Fig. 5a, multiple free-standing films of heterostructure material are stacked up to form a thick detector, as discussed in the previous section, a single optical detector being attached to (or grown on) one or both sides of the stack.

The second, more sophisticated level of integration, shown in Fig. 5b, implies multiple stacking of bulk scintillators, each having individual photo-receiver. The optical signals from the multiple photo-receivers are electronically processed separately. We are no longer concerned with internal reflections because we are stacking complete systems that share the penetrating high-energy radiation but not the scintillating photons. Such an approach allows one to both increase the total device thickness and obtain an information on the penetration depth of the ionizing radiation. Furthermore, simultaneous signal registration in several detectors in the stack allows one to extract spectral characteristics free from complications associated with Compton scattering processes, see, e.g., Chap. 10 in [1]. One should not underestimate, however, the difficulty inherent in manipulating nanosecond pulses. In order to deliver information in this raw analog form, one needs sophisticated low-capacitance interconnects that can handle short pulses.

This difficulty is removed in the highest level of integration, which is achieved when the electrical signals from each individual detector slab are electronically processed in each slab and converted to a digital information. In this case, every detector slab in the stack represents an opto-electronic chip, delivering the information in a digital, noise-free format. The detailed discussion of the electronic architecture of such a 3D integrated system will be presented elsewhere. Here it sufficient to emphasize that each integrated chip must report not a 1 ns pulse, but a digital signal carrying the required information – where in the stack the photo-multiplication occurred, the time of the event, its amplitude, and the lateral coordinates of the event. Such an integrated system offers enormous advantages for definite error-free identification and characterization of high-energy radiation, as well as accurate determination of the direction to source.

## 7. Discussion

The proposed semiconductor high-energy detector possesses unique properties: it operates as a solid-state scintillator, with high detection efficiency and a response time in the nanosecond range. The relatively low energy gap, and therefore low energy per single photon (4 eV for InP versus 25 eV for NaI) gives the absolute photon yield of ~ 250,000 photons/MeV, versus ~ 40,000 photons/MeV for NaI.

It is instructive to compare the absolute scintillator efficiencies, defined [1] as $\eta = (E_\gamma / 3E_G) \times (h\nu / E_\gamma)$, where $E_\gamma$ is the particle energy, $3E_G$ the average energy to create an electron-hole pair, and $h\nu$ the energy of scintillating photons. In thallium-activated NaI electron-hole pairs are produced across the band-gap of over 7 eV while the scintillating photons are produced at $h\nu$ of only 3 eV, whence the best $\eta$ available in NaI scintillators is about 12%. In contrast, the effective value of $3E_G$ in both InP and GaAs is 4.1 eV [14,15] and the $h\nu$ of scintillating photons is close to $E_G$. We can expect therefore an improvement in $\eta$ by a factor of 7/3 over thallium-activated NaI.

The direct-gap semiconductor scintillator is expected to provide an energy resolution similar to that of Si and Ge detectors. The excellent spectral resolution is expected from similarity of the semiconductors. Thus, statistical properties of electron-hole pair creation in GaAs should close resemble similar process in germanium. Special effort is required, however, both experimental and theoretical, to quantify the statistics of energy resolution, in particular the device Fano factor [16, 17].

The device does not need high voltages[2] or low temperatures for operation and is robust against the hostile environment, e.g. it can operate in high magnetic fields. The preferred embodiment comprises InP, because of the higher atomic number of In, but other appropriately doped direct gap semiconductors, such as GaAs, can be used too and may be advantageous from the point of view of cost and the ease of 3D integration.

---

[2] The scintillator itself requires no electrical bias. Small voltage must be applied to operate an integrated detector of light produced by the scintillator.

The proposed detector can be used both as an inexpensive monitor of the radiation environment, like a Geiger counter, and as a high-resolution radiation spectrometer. The more sophisticated versions of the proposed device, involving either epitaxial growth and stacking of free-standing structures or 3D integration, offer unprecedented quality of particle identification in spectral characterization of the high-energy radiation.

Among the many attractive features of the proposed detector, we should single out its exceptionally fast response. We note that the prompt response of our detector is *not* owing to the quenching of the radiative transitions by non-radiative processes and, therefore, is not accompanied by any degradation of brightness. Our fast response is based on the fact that radiative transitions are themselves fast in a direct-gap semiconductor at a sufficiently high majority-carrier concentration. This property has not been utilized before in a high-energy radiation detector because fast transition rates are normally associated with short mean-free path of the resultant scintillating radiation. *Breaking this association is the essence of our invention*. Our detector is simultaneously fast and bright. The other essential part of our invention is that it enables novel system architectures, discussed in Sect. 6, where each relatively thin semiconductor slab is supplied with its own photo-detector system. Such systems can then be stacked up without limit, thus increasing the active detector volume to accommodate large absorption length of high-energy radiation.

Another attractive feature of the integrated detector stack is that it enables 3D pixellation of the scintillator response. As will be described elsewhere, this in turn enables signal processing with tantalizing capabilities, such as the accurate determination of the direction to source of high-energy radiation. It also offers new means for spectral characterization of incident particles – in addition to and independent of the conventional statistical analysis [1] of the total yield per particle.

**Acknowledgement**

We would like to thank Prof. Gregory Belenky for a useful discussion of nonradiative recombination.

**References**

[1]     G.F. Knoll, *Radiation Detection and Measurement*, 3rd edition, John Wiley & Sons, Inc. (2000).

[2]     W. Lehman, "Edge emission of n-type conducting ZnO and CdS", *Solid-State Electron.* **9**, pp. 1107-1110 (1966).

[3]     S.E. Derenzo, M.J. Weber, E. Bourret-Courchesne, M.K. Klintenberg, "The quest for the ideal semiconductor scintillator", *Nucl. Instr. and Meth. in Phys. Res*. **A 505**, pp. 111-117 (2003).

[4]     S.E. Derenzo, E.Bourret-Courchesne, M.J.Weber, M.K.Klintenberg, "Scintillation studies of CdS(In): effects of codoping strategies", *Nucl. Instr. and Meth. in Phys. Res*. **A 537**, pp. 261-265 (2005).

[5]     S. E. Derenzo, private communication.

[6]     E. Burstein, "Anomalous Optical Absorption Limit in InSb", *Phys. Rev.* **93,** pp. 632-633 (1954).

[7]     G. P. Agrawal and N. K. Dutta, *Semiconductor lasers*, 2nd edition, Van Nostrand, New York (1993).

[8]     W.P. Dumke, M.R. Lorenz, and G.D. Pettit, "Intra- and Interband Free-carrier Absorption and the Fundamental Absorption Edge in *n*-Type InP", *Phys. Rev.* **B 1**, pp. 4668-4673 (1970).

[9]     I. Balslev, "Optical Absorption due to Inter-Conduction-Minimum Transitions in Gallium Arsenide", *Phys. Rev.* **173,** pp. 762-766 (1968).

[10]    R. Windish, C. Rooman, B. Dutta, A. Knobloch, G. Borghs, G.H. Döhler, P. Heremans, "Light-Extraction Mechanisms in High-Efficiency Surface-Textured Light-Emitting Diodes", *IEEE J. Select. Topics in Quantum Electronics* **8**, pp. 248-255 (2002).

[11]    V. Zabelin, D.A. Zakheim, S.A. Gurevich, "Efficiency Improvement of AlGaInN LEDs Advanced by Ray-Tracing Analysis", *IEEE J. Quantum Electronics* **QE-40,** pp. 1675-1686 (2004).

[12]    K. Christiansen, M. Luenenbuerger, B. Schineller, M. Heuken, and H. Juergensen, "Advances in MOCVD technology for research, development and mass production of compound semiconductor devices", *Opto-Electronics Review,* **10**(4), pp, 237-242 (2002).

[13]    H. Morkoç, "Comprehensive characterization of hydride VPE grown GaN layers and templates", *Mat. Sci. and Eng.,* **R33**, pp. 135-207 (2001).

[14]    C.A. Klein, "Bandgap dependence and related features of radiation ionization energies in semiconductors", *J. Appl. Phys.* **39**, pp. 2029-2038 (1968).

[15]    D.S. McGregor and H. Hermon, "Room-temperature compound semiconductor radiation detectors", *Nucl. Instr. and Meth. in Phys. Res*. **A 395**, pp. 101-124 (1997).

[16]    U. Fano, "Ionizing yield of radiations. II. The fluctuations of the number of ions", *Phys. Rev.* **72,** pp. 26-29 (1947).

[17]    T. Papp, M.C. Lépy, J. Plagnard, G. Kalinka, E. Papp-Szabó, "A new approach to the determination of the Fano factor for semiconductor detectors", *X-ray Spectrometry* **34**, pp. 106-111 (2005).

**Figure Captions**

**Figure 1. Burstein shift.**

Absorption of photons $h\nu$, emitted by recombining electrons and hole in the semiconductor, is largely suppressed by the absence of vacant electron states in the conduction band under the Fermi level in heavily doped semiconductor.

**Figure 2. Optical absorption processes in a direct-gap semiconductor.**

1. Interband absorption; 2. Intraband free-carrier absorption; 3. Free-carrier absorption between different conduction bands.

**Figure 3. Concentration dependence of the absorption length.**

Calculated absorption length $\lambda$ in InP at optical wavelength of 0.92 μm is plotted as function of doping concentration $N_D$ for different temperatures. The free-carrier contribution, shown separately, has been obtained from the data of Dumke et al [8].

**Figure 4. Heterostructure detector based on bandgap engineering.**

The heterostructure comprises two alternating materials, exemplarily InP and InGaAs ternary alloy, lattice matched to InP, forming narrow wells where light emission occurs.

**Figure 5. Three levels of integration.**

(a) Stacking heterostructure films using a high-index fluid. Single photo-receiver is used for signal registration.

(b) Stacking detector slabs with individual photo-receivers. Analog signals (electrical pulses) from each detector are delivered to a central processing unit.

(c) Complete detector integration system. The electrical signals are digitally processed at each detector slab, and information is transmitted to the central unit in digital form.

# Figures

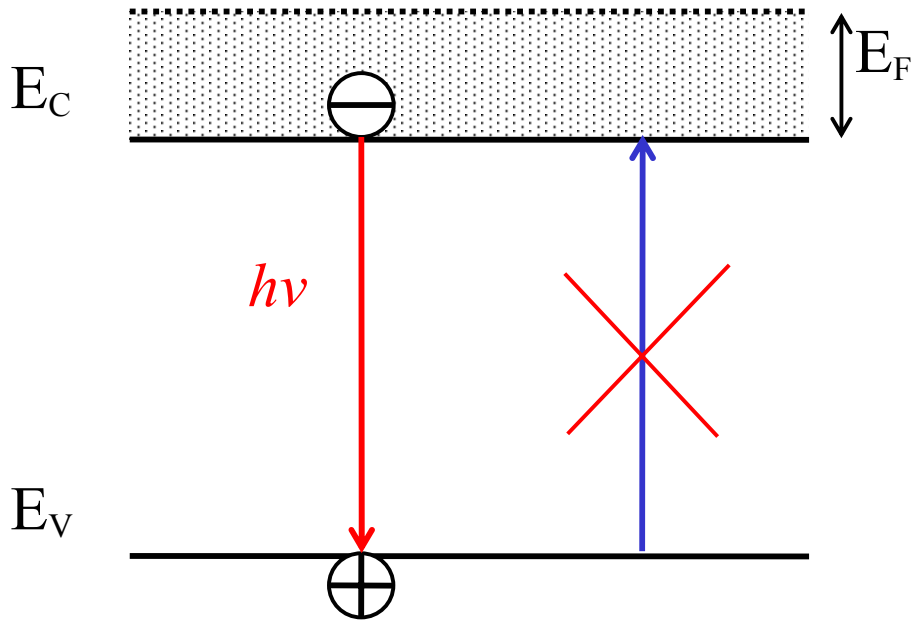

**Figure 1. Burstein shift.**

Absorption of photons *hv*, emitted by recombining electrons and hole in the semiconductor, is largely suppressed by the absence of vacant electron states in the conduction band under the Fermi level in heavily doped semiconductor.

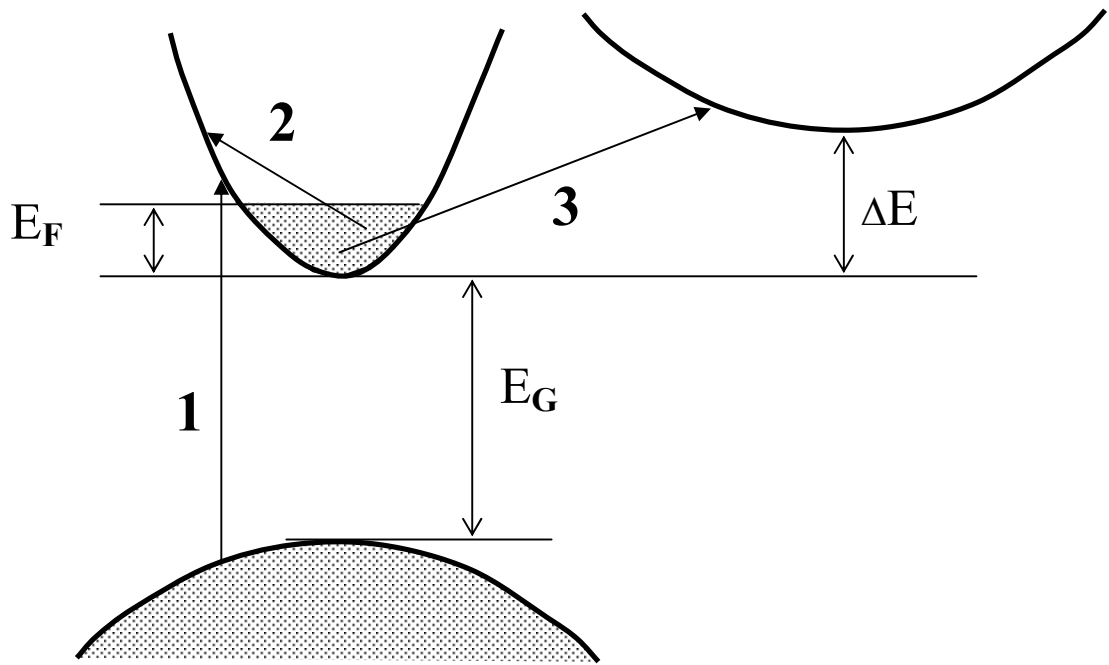

**Figure 2. Optical absorption processes in a direct-gap semiconductor.**

1. Interband absorption; 2. Intraband free-carrier absorption; 3. Free-carrier absorption between different conduction bands.

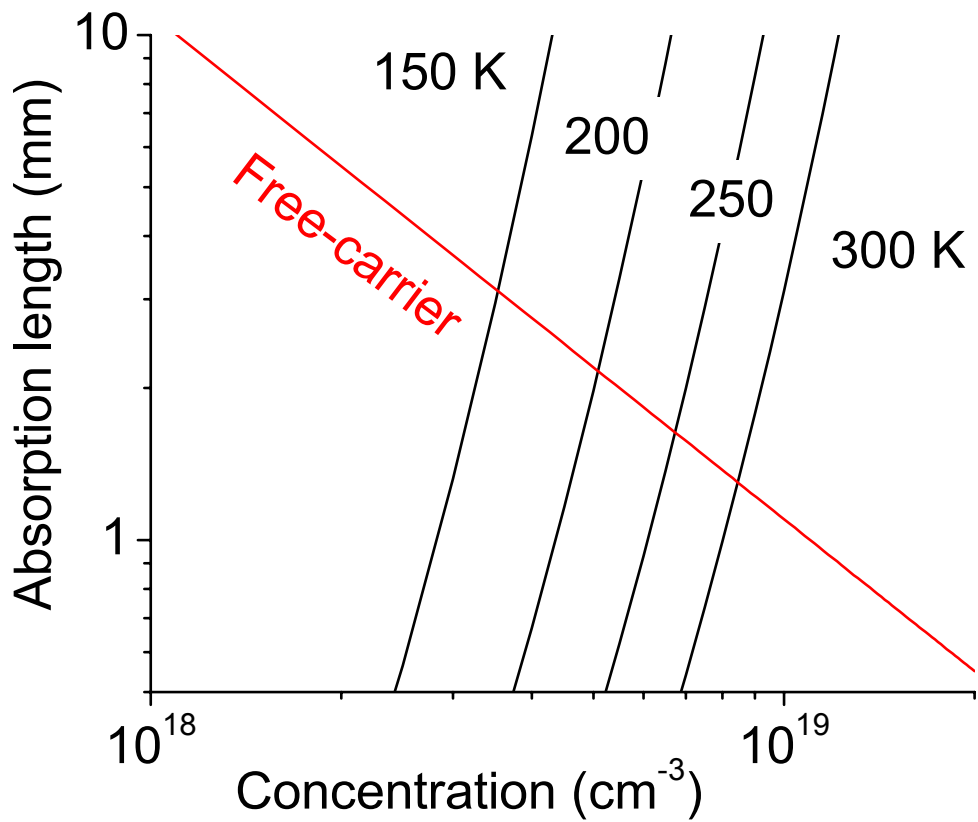

**Figure 3. Concentration dependence of the absorption length.**

Calculated absorption length $\lambda$ in InP at optical wavelength of 0.92 μm is plotted as function of doping concentration $N_D$ for different temperatures. The free-carrier contribution, shown separately, has been obtained from the data of Dumke et al [8].

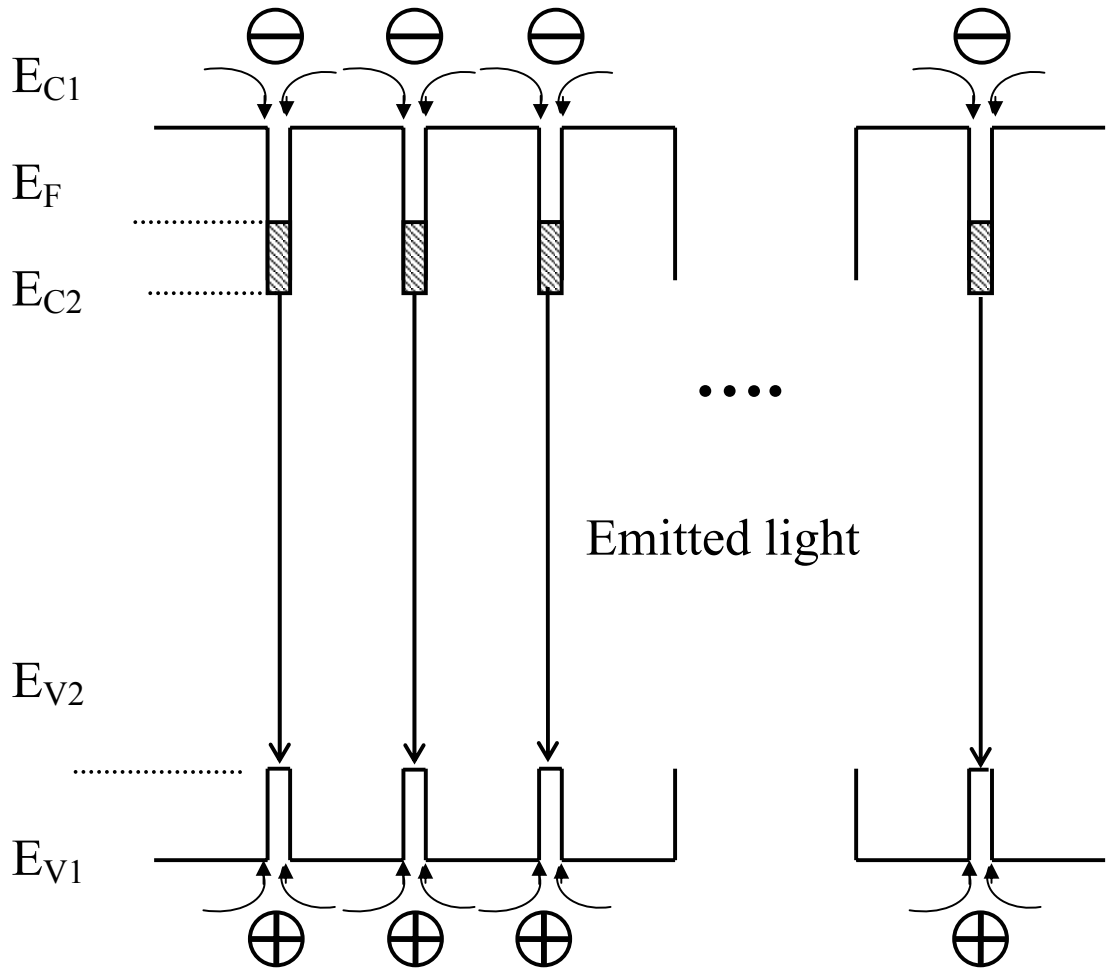

**Figure 4. Heterostructure detector based on bandgap engineering.**

The heterostructure comprises two alternating materials, exemplarily InP and InGaAs ternary alloy, lattice matched to InP, forming narrow wells where light emission occurs.

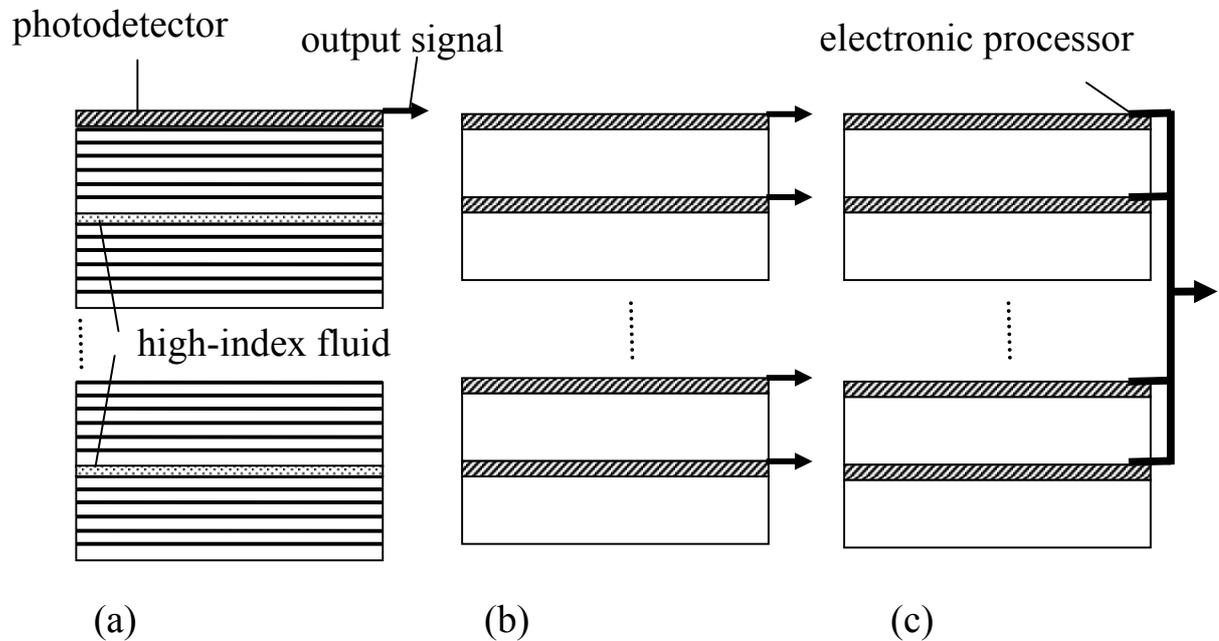

**Figure 5. Three levels of integration.**

(a) Stacking heterostructure films using a high-index fluid. Single photo-receiver is used for signal registration.

(b) Stacking detector slabs with individual photo-receivers. Analog signals (electrical pulses) from each detector are delivered to a central processing unit.

(c) Complete detector integration system. The electrical signals are digitally processed at each detector slab, and information is transmitted to the central unit in digital form.